# Periodic domain inversion in single crystal barium titanate-on-insulator thin film

*Pragati Aashna[1], Hong-Lin Lin[1], Yu Cao[1*], Yuhui Yin[1], Yuan Gao[1], Sakthi Sanjeev Mohanraj[3], Di Zhu[2,3], and Aaron Danner[1*]*

[1] Department of Electrical and Computer Engineering, National University of Singapore, 117583, Singapore

[2] Department of Material Science and Engineering, National University of Singapore, 119077, Singapore

[3]Institute of Materials Research and Engineering, Agency for Science, Technology and Research, 138634, Singapore

[*]*elecao@nus.edu.sg*, *adanner@nus.edu.sg*

We report experimentally achieving first-ever electric field periodic poling of single crystal barium titanate (BTO, or $BaTiO_3$) thin film on insulator. Owing to the outstanding optical nonlinearities of BTO, this result is a key step towards achieving quasi-phase-matching in BTO. We first grow the BTO thin film on a dysprosium scandate substrate using pulsed laser deposition with a thin layer of strontium ruthenate later serving as the bottom electrode for poling. We present characterization of the BTO thin film using x-ray diffraction and piezo-response force microscopy to clearly demonstrate single crystal, single domain growth of the film which enables the desired periodic poling. To investigate the poling quality, we apply both non-destructive piezo force response microscopy and destructive etching-assisted scanning electron microscopy and we show that high quality, uniform and intransient poling with 50 % duty cycle and periods ranging from 2 µm to 10 µm is achieved. The successful realization of periodic poling in BTO thin film unlocks the potential for highly efficient nonlinear processes under quasi-phase-matching that seemed far-fetched with prior polycrystalline BTO thin films which predominantly relied on efficiency-limited random or non-phase matching conditions and is a key step towards integration of BTO photonic devices.

## 1. Introduction

Barium Titanate (BTO) was one of the first ferroelectric materials to be discovered and it still maintains prominence owing to its chemical and mechanical stability, ferroelectric properties at or above room temperature, and strong linear and nonlinear electro-optic coefficients. It has been hypothesized for many decades that these superior properties will eventually lead to multi-faceted applications in optical communications, sensors and memory devices,[1-5] but concrete achievements have largely remained elusive due to challenges in crystal growth and fabrication. Recently, thin film BTO-on-insulator has garnered increasing attention, after it was reported that pulsed laser deposition is able to create high quality single crystal BTO thin films on a substrate with a lower refractive index, enabling planar photonic integration.[6-13] Apart from having electro-optic coefficients much larger than those of lithium niobate (LN),[9,10] thin film BTO has been shown to have a large transparency window in the visible and infrared ranges, nonlinear properties which can be tuned under strain, and the possibility of integration with silicon photonics.[6-12] Mode confinement in thin film BTO through rib waveguide structures has also been reported.[12-20]

Ferroelectric materials like BTO possess a net dipole moment whose interaction with light gives rise to nonlinear optical processes like second harmonic generation (SHG), sum frequency generation, difference frequency generation, etc., potentially allowing applications such as parametric amplifiers, conversion of photons in the UV, IR or THz bands, frequency combs, creation of efficient entangled photon pairs for quantum optical applications, etc.[21-24]. However, these three-wave-mixing-based nonlinear optical processes are naturally inefficient due to inherent dispersion which results in phase mismatches between the interacting waves. One of the most versatile techniques to achieve phase matching is quasi-phase-matching (QPM) achieved via electric field poling of a ferroelectric material by introducing periodic perturbations in its nonlinear susceptibility.[21-23] QPM-based nonlinear processes through electric field poling have been extensively reported in both x and z-cut thin-film LN (TFLN) platforms because of the availability of good quality commercially available single crystal c-domain TFLN wafers.[24-27] However, until now, periodic poling for QPM has not been achieved in BTO thin films due to the unavailability of high-quality single crystal BTO thin films. Though it has been reported in bulk crystals, to the best of our knowledge, only one such report in bulk can be found in literature.[28]

To achieve vertical optical index confinement, low index substrates such as MgO have typically been used in the past to grow thin film BTO, but the lattice mismatch results in polycrystalline BTO containing multiple domains.[12-20] Although such thin film BTO is not single crystal, nonlinear processes such as second harmonic generation have been extensively explored and demonstrate strong second-order nonlinearities in the visible to IR ranges.[29-34]

Due to the presence of both 'a' and 'c' crystal orientations in these films, achieving periodic poling poses significant challenges. The coexistence of multiple domains introduces complexity, as each domain may respond differently to the applied poling field, making uniform and precise poling more difficult to attain. As a result, all nonlinear processes previously reported in BTO thin films are either non-phase matched, or they satisfy random quasi-phase matching conditions (the averaging of nonlinear polarization over several randomly oriented domains) with only limited conversion efficiencies due to the phase matching being only approximately satisfied over small distances. Statistical variations in BTO grains also lead to reduction in conversion efficiencies compared to monocrystalline BTO, since it is impossible for any design to target the relevant nonlinear susceptibility component.[29-34]

Recently, we have reported very high-quality single-crystal single-domain BTO thin film on insulator, grown using pulsed laser deposition on a dysprosium scandate (DSO, or $DyScO_3$) substrate. Notably, DSO has a lower refractive index than BTO, providing an ideal platform to fabricate optical structures with excellent vertical mode confinement.[35-37] The BTO thin films are aligned with a vertical c-axis which facilitates the required electrode placement during poling.

Here we present, to the best of our knowledge, the first-ever experimental demonstration of periodic poling in thin film single crystal BTO on insulator. To achieve poling, we apply an electric field larger than the coercive field strength of BTO along the c-axis, to permanently invert the spontaneous polarization direction. With appropriate lithographically defined electrode designs, periodic poling of thin film BTO with any desired period and duty cycle can be achieved using the correct electric field pulse shape. We incorporate two methodologies to verify the quality and extent of poling. The first approach is piezo force microscopy (PFM), and the other technique is selective etching, where the poled sample is wet-etched and the differing etching rates of antiparallel domains reveals the poling pattern which can then be imaged using scanning electron microscopy (SEM). We have also verified that poling is permanent by re-measuring samples over several weeks.

Our research offers the potential of integrating BTO photonic devices, previously hindered by challenges in achieving efficient nonlinear processes crucial for diverse optical applications, ranging from frequency comb generation to entangled photon pair production. This positions our findings as a promising alternative to traditional TFLN devices.

## 2. Results and Discussions

### 2.1 BTO thin film growth and characterization

BTO (c-axis) thin film is grown on a DSO substrate by pulsed laser deposition with a KrF laser (248 nm wavelength). However, since the spontaneous polarization of BTO is out-of-plane, a vertical electrode geometry will later be needed to apply a poling field; therefore, we first grow a thin

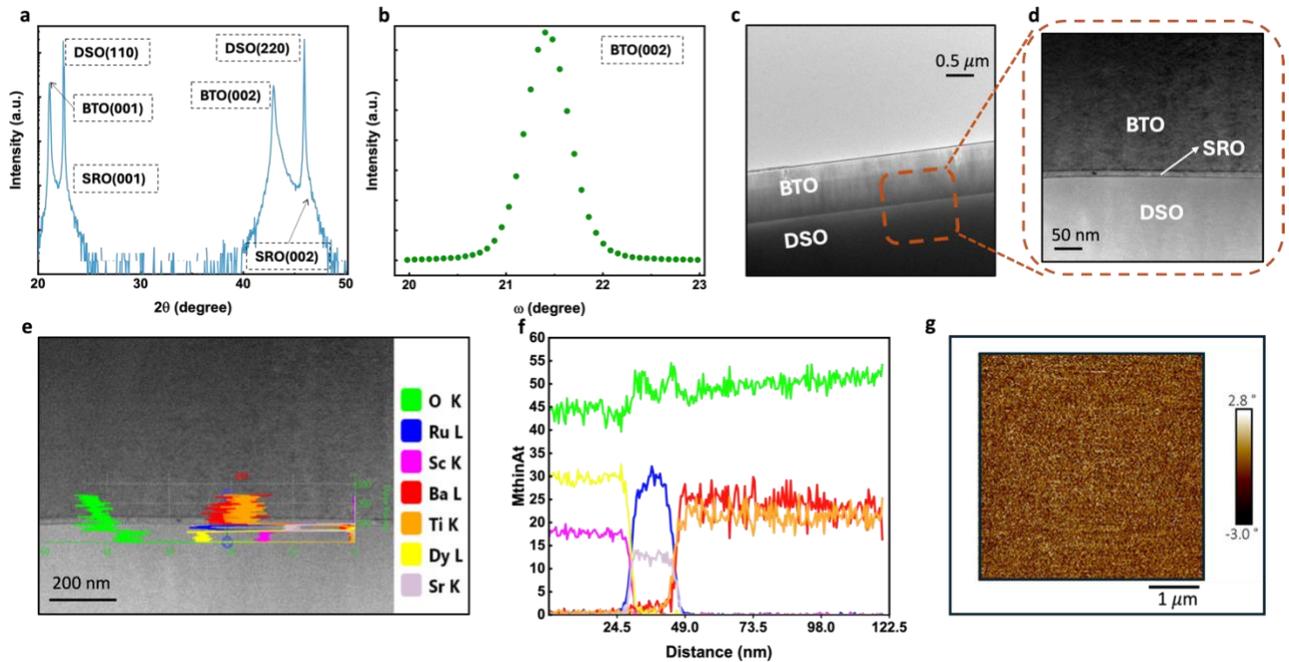

**Figure 1** (a) X-ray diffraction (XRD) spectra for BTO/SRO/DSO, XRD 2θ-ω coupled scan showing BTO is single crystalline with c-axis oriented out-of-plane. (b) XRD rocking curve at the BTO (002) peak with the full width at half maximum (FWHM) of 0.53° shows high crystal quality for BTO (c) SEM image of the BTO/SRO/DSO cross section. The SRO is only 15 nm thick, so it is not visible (d) shows the enlarged view of the cross section to clearly demonstrate the three layers (e) Energy dispersive X-ray analysis (EDXA) for the chemical characterization of the sample (f) showing the corresponding variation in elemental components along the cross-section of the structure (g) PFM phase image of the film surface showing well-aligned domain orientations layer of 15 nm strontium ruthenate (SRO, or $SrRuO_3$) as bottom electrode. SRO has a similar crystal lattice to BTO and DSO and is conductive. SRO is grown at 600 °C under 100 mTorr oxygen gas pressure with a laser pulse energy density of 2 J·cm$^{-2}$. Then 750 nm BTO thin film is grown epitaxially on top of the SRO layer at 650 °C under 10 mTorr with a laser pulse energy density of 1.5 J·cm$^{-2}$, followed by cooling at 100 Torr oxygen gas pressure. The growth rates of thin films are evaluated by step profiler and X-ray reflectometry, and film thickness is controlled by the growth time. The crystallinity of the thin films is characterized by X-ray diffraction (XRD), as shown in **Figure 1**(a) & (b). We show in **Figure 1** (c) the SEM image of the cross-section of the grown BTO/SRO/DSO sample. Since SRO is only 15 nm thick, we also show an enlarged image of the cross section in **Figure 1**(d), clearly showing all the three layers. We also perform the EDXA for chemical characterization of the hence grown sample in **Figure 1** (e) and (f).

XRD characterization reveals that the BTO is well-oriented along the c-axis. However, XRD does not provide information about the potential presence of antiparallel domains exhibiting out-of-plane spontaneous polarizations. Ensuring the absence of such domains is crucial, as they could interfere with the efficient phase matching required under the QPM process. To address this, we conduct PFM imaging of the sample prior to poling, examining various regions of the film surface. A representative image is presented in **Figure 1**(g), where the phase image confirms the absence of antiparallel domains, verifying the alignment of all domains.

## 2.2 Sample preparation and experiment

Given that our thin film has a c-axis orientation, it is required to perform electric field poling of the BTO thin film by fabricating the periodic electrode on the top of the film, and then applying the field along the -z-direction with SRO acting as the ground.

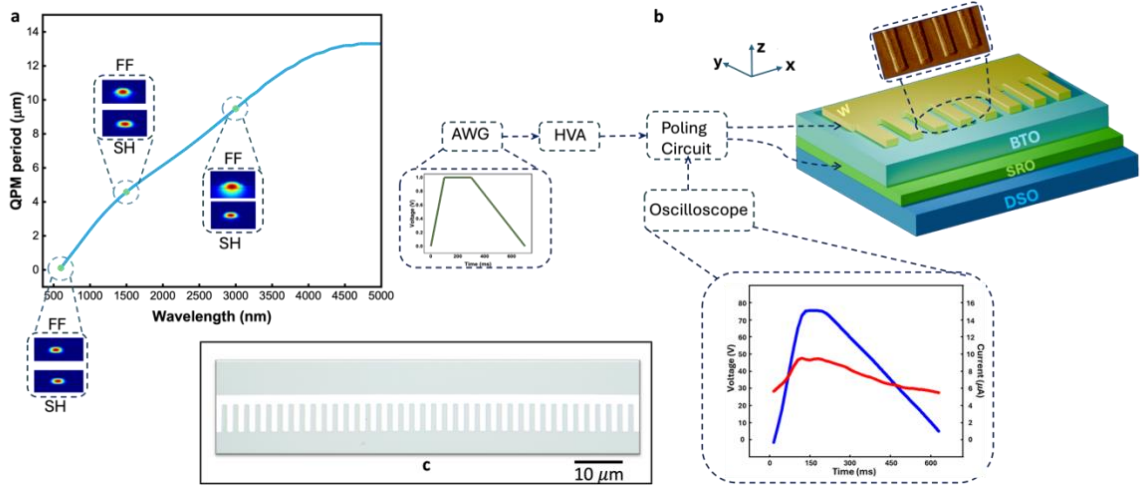

**Figure 2 (a)** Variation of QPM period for SHG with the fundamental wavelength, insets show the modal profiles for the fundamental TE modes for frequencies at 600 nm, 1500 nm and 3000 nm and their respective SH profiles. **(b)** Poling set up: An arbitrary waveform generator (AWG) is used to generate a trapezoid-like waveform which is amplified by a high voltage amplifier (HVA) and applied across the BTO thin film via a poling circuit. Thin film BTO on DSO with a finger electrode pattern on top (inset showing the AFM image of the electrodes patterned on our sample) and a ground SRO electrode beneath the BTO film are shown. The input waveform from the AWG and that observed at the oscilloscope (blue waveform correspond to applied field and red refers to the resulting poling current across a load) are also shown **(c)** The optical microscope image of the patterned electrodes.

Since the transparency window of BTO thin film spans from visible to IR wavelengths, the QPM period for later applications like SHG would vary accordingly. Since nonlinear processes must ultimately be implemented in waveguides, we take actual rib waveguide parameters from previously fabricated devices on BTO thin film and simulate the required QPM period for SHG as a function of fundamental wavelength (from 600 nm to 5 μm) as shown in **Figure 2** (a).[13] We observe that the waveguide reaches cut-off at 3.5 μm, which gives us an estimate for practical poling periods, which should ideally fall within the range of 2-10 $\mu m$. We thus target these periods for poling as they are likely to be the most useful in applications.

To form the top poling electrodes, we deposit tungsten (W) on top of the BTO thin film and pattern the electrodes using electron-beam-lithography (EBL-Raith). We then dry etch the W using inductively coupled plasma reactive ion etching (ICP-RIE) to obtain the final finger-like electrode pattern shown in **Figure** 2 (c). The length of the electrodes in the x-direction ranges from 500 μm to 5 mm.

The poling experiments are carried out at room temperature using the set up shown in **Figure 2** (b). Though the arbitrary wavefunction generator serves as the input, we externally generate the waveform and adjust its parameters through Python programming with a computer. The waveform is then amplified using a programmable high voltage amplifier and is applied to our sample using an appropriate poling circuit. We use an oscilloscope to visualize the poling waveform and calculate the resulting poling current across a load as shown in **Figure 2** (b).

### 2.3 Poling parameters optimization

When an electric field is applied during poling of any ferroelectric material, nucleation of domains occurs first. As the voltage continues to rise, both vertical and lateral expansion of inverted domains occur until they coalesce. Key indicators of high-quality periodic poling include control over poling period, duty cycle, homogeneity of poling over length, and the achieved depth of inverted domains. These factors are instrumental in facilitating efficient nonlinear processes.

To optimize these parameters and ensure superior quality poling, we conduct optimization experiments on several samples equipped with electrodes between 500 $\mu m$ to 5 mm in length. We

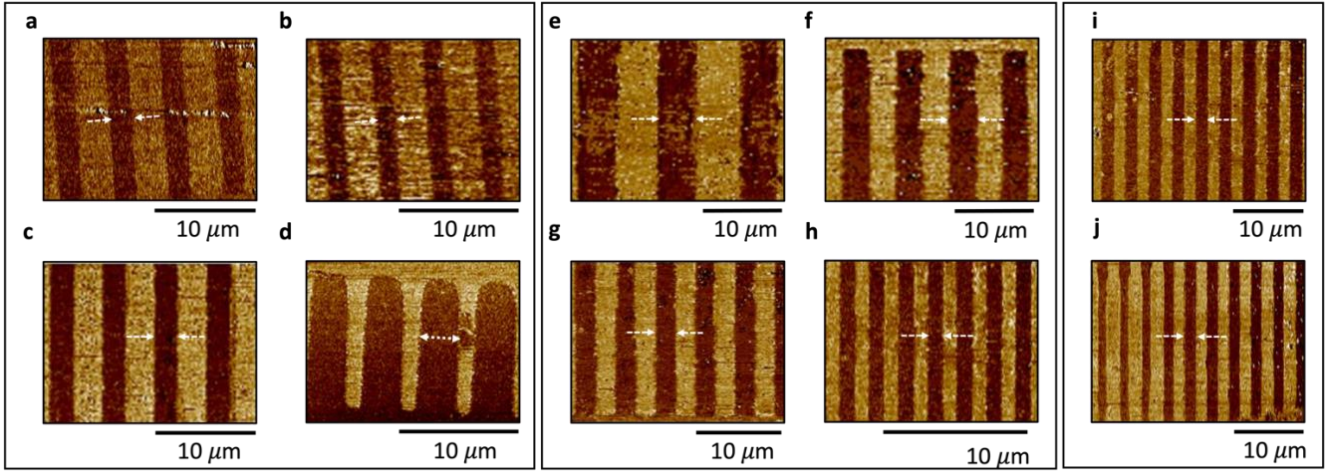

**Figure 3.** (a)-(c) PFM phase images corresponding to under-poling and (d) over-poling. (a) & (b) correspond to under-poling resulting from applying a voltage less than the voltage required for complete domain inversion (c) The domain inversion is complete as evident from the image contrast, but the duty cycle is only 45% (d) Over-poling which results from domain spreading beyond electrode fingers. (e)-(i) PFM images for inverted domains with 50% duty cycle with different periods: (e) 10 μm (f) 7 μm (g) 5 μm (h) 2 μm (i) 3 μm domain with complete electrode finger length imaged (j) Aperiodic poling showing domain inversion corresponding to electrodes with varying widths and period

optimize the following parameters to achieve high-fidelity poling:

*Poling voltage:* From the ferroelectric hysteresis curve, the BTO coercive field strength is found to be around 50 kV/mm. Given our film thickness of 750 nm and the bottom electrode lying just below the BTO, we initiated the poling process by incrementally increasing the poling voltage from 45 V (slightly surpassing the coercive voltage, obtained from hysteresis analysis of BTO film) in steps of 5 V. Domain nucleation starts with the application of the field, but the inversion does not take place until 55 V. Though poling initiates at 60 V, high-quality domains start forming only beyond 70 V.

*Pulse duration:* The pulse duration is contingent upon the dimensions of the region to be poled and also the voltage applied which generates the poling current. There is no guaranteed technique to determine this, so we vary the poling pulse duration (time for which the voltage remains constant) from 10 ms to 500 ms on different electrodes and observed the results for optimization.

*Number of pulses*: To optimize the number of pulses, we explored single pulse waveforms up to nearly two dozen pulses. We noticed that the number of pulses had little impact on poling outcome. Consequently, we applied either 10 or 20 pulses.

It may be noted here that we used a conventional waveform, widely accepted for poling, consisting of a steady increase in voltage, followed by the application of a constant voltage and finally, a gradual decrease till zero, as shown in **Figure 2** (b).

Following the optimization of poling parameters on test samples, we proceed to fabricate 5 mm long periodic electrodes with periods ranging from 2 to 10 μm. Various combinations of pulses were then applied to different electrodes based on the optimized parameters. From the optimization experiments, we observed that for applied voltages within the range of 90-110 V, the optimized pulse parameters include a total pulse duration of 700 ms, with a rise time of 50 ms, a poling time of 300 ms, and a ramp-down time of 350 ms. These parameters result in periodic poling with 50% duty cycle, regardless of whether 10 or 20 pulses were applied. We intuitively followed a pattern of giving more pulses at lower voltages and vice versa. After poling, we removed the W electrodes from the surface and used piezo-response force microscopy (PFM) to image the poled regions.

### 2.4 PFM imaging results

We utilize the PFM imaging technique to assess and evaluate the quality of poling in relation to parameters such as period, duty cycle, and homogeneity. With PFM, we visualize antiparallel (180°) ferroelectric domain structures by detecting sample contraction and expansion through an appropriate probe. This process, operating under the converse piezoelectric effect, provides both amplitude and phase data for the measured deflections, aiding in comprehensive analysis and interpretation. Here, we introduce three types of data that are pertinent to any periodic poling procedure.

First, we report under-poling, where duty cycle of the poled region is less than 50% and the phase shift between the adjacent domains is not prominent, as shown in **Figure 3** (a) and (b). This occurs when either the applied peak voltage or the poling pulse duration is insufficient for lateral and vertical expansion of the domains but sufficient to start the nucleation of domains, therefore negligible phase contrast between the adjacent domains is detected. For these electrodes, we applied three pulses of 60 V peak voltage and 500 ms pulse duration. When we increased the voltage to 75 V keeping all the other parameters the same, we did observe domain inversion, but with a duty cycle of only 45 % as shown in **Figure 3** (c).

Second, we report over-poling, or expansion of domains laterally such that the duty cycle is more than 50 %. This happens when either, or both, applied voltage and pulse duration are high enough to facilitate lateral expansion of the domains beyond the width of electrode fingers. We can see from **Figure 3** (d) that the poled domains have a duty cycle of around 70 % as we applied a voltage of 120 V with pulse duration of 2100 ms. Both under poling and over poling are undesirable for practical applications.

The third case corroborates periodic poling with a 50% duty cycle which is desirable for maximum conversion efficiency of most nonlinear processes. Using our estimated simulation data, we try to achieve this for periods ranging from 2 to 10 $\mu$m. We use the optimized parameters for the input waveform with peak applied voltage in the range 90-110 V and pulse duration of 700 ms. We apply 10 or 20 pulses. PFM images of inverted domains are shown in **Figure 3** (e)-(h). We can see from the figure that 50% duty cycle can be realized for all depicted periods. Additionally, to confirm an even distribution of poling along the entire structure length, we image the ends and the center of the poled region, and we observed that the poling is uniform along the length with no noticeable change in the duty cycle. For PFM, the sensitivity of the probes lies in the $\mu$m range and since our film is 750 nm thick, we can conclude from the PFM images that the poling is uniform across the thickness of the film as well. Furthermore, we also varied the length of the electrode fingers (along y-direction) from 10 $\mu$m to 25 $\mu$m and **Figure 3** (i) shows the full length of 25 $\mu$m teeth are poled uniformly.

For several practical applications, it is imperative to vary the poling period along the length of the device; for example, for wideband frequency conversion, chirped poling is required.[38] Therefore, it is pertinent for us to demonstrate this case for which we chose a range of periods along the length of a device and verified the result, as shown in **Figure 3** (j), that the poling imitates the defined pattern, hence exhibiting the robustness of poling.

## 2.5 Selective etching and SEM

For z-cut LN, selective etching can be considered as the standard technique and final check on poling quality vis-à-vis the key poling parameters. The same technique can be extended to BTO thin film as well.

To expose the domains, we etch our poled samples in 0.03 % HF solution for 2 minutes 30 seconds and we examine the poling quality under SEM. It can be seen from **Figure 4** (a)-(b) that the poling is uniform with 50 % duty cycle. We image the whole 5 mm sample for each poled electrode, and we observed that the poling is homogeneous along the full length. Therefore, we can say with certainty that the poling has both lateral and vertical homogeneity. Moreover, we also show using a tilted SEM image in **Figure 4** (c) that the domain walls are free from imperfections, which may not be conferred from the PFM images, and henceforth we provide a holistic set of data to substantiate our high-fidelity poling results.

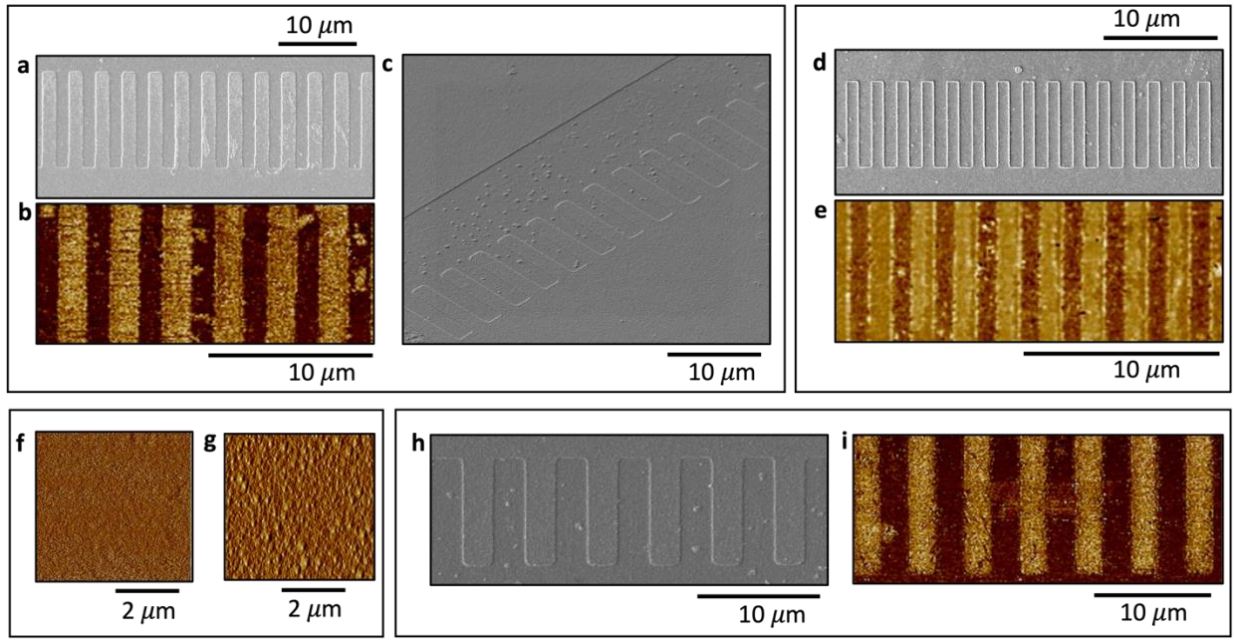

**Figure 4.** (a-b) Scanning electron microscopy (SEM) images and corresponding PFM images of poling pattern revealed after selective HF etching for 150 seconds showing good uniformity in poling along the length and depth (c) Tilted SEM image to show the smooth sidewalls of the poled region, which further endorses the high-poling quality. (d)-(e) Scanning electron microscopy (SEM) and corresponding PFM images of poling pattern revealed after selective HF etching for 300 seconds, showing good uniformity in poling along the length and depth (f)-(g) Etching results into roughness which exacerbates the PFM image quality, and we conform the roughness through the AFM images, first one refers to before etching and second after etching (h)-(i) SEM and PFM phase image of inverted domains after 3.5 months of poling. The image is achieved after selective etching of the sample in HF for 120 seconds.

To check the vertical uniformity further, we put the sample in the HF solution for 5 minutes, hence etching twice as long before. We can see from **Figure 4** (d)-(e) that the inverted domains are still consistent and uniform across the film thickness while maintaining the periodicity, duty cycle as well as homogeneity along the length.

For both the cases, after etching, we also perform PFM imaging (bottom images in 4 (b) and (e) on these samples to further validate the existence of inverted domains across the film thickness and show that they agree well with the SEM results. It may be observed that due to wet etching, the surface roughness of the film increases which exacerbates the poling image quality through PFM. We verify that by showing the AFM images for roughness of the film surface before and after the etching in **Figure 4** (f) and (g).

In numerous ferroelectrics, the process of poling is transient and tends to diminish over time, thereby restricting their utility in photonic circuits or optical memory devices. For our poled BTO samples, we perform PFM imaging weekly on the same sample for 3 weeks and observed that the poling stays without losing its quality. However, for a test sample which we poled first, when we imaged after 3.5 months, we did not observe very good contrast in PFM phase images of the antiparallel domains. But when we etched the sample in HF for 2 minutes and again conducted PFM on it, we could see z topography of the poled structure due to the selective etching even under the optical microscope. We then obtained the PFM and SEM images to validate the result as shown in **Figure 4** (h)-(i) which implies that the disappearance might be attributed to some surface effect.

The PFM images obtained after poling, followed by the subsequent HF etching, provide compelling evidence that periodic poling has been effectively implemented in our BTO on insulator thin film.

## 3. Conclusion

In conclusion, we have reported the first ever electric field poling of thin film BTO on DSO substrate and we show that high fidelity poling can be obtained on our samples. Poling is performed by applying high voltage pulses to the patterned electrodes on the top of the thin film. We substantiate our results using PFM imaging that shows excellent contrast between the anti-parallel domains. We also apply the versatile and most acceptable technique of selective etching to investigate the poling quality. We present convincing evidence to show that the inverted domains are uniform and homogeneous along the full length of the device and through the whole film thickness with periods as small as 2 $\mu$m and 50% duty cycle. The achieved periodic poling will be further extended to study SHG under QPM in thin film BTO waveguides in our subsequent work. This will unlock the full potential of integrated optical devices based on thin film BTO which hitherto were limited due to inefficient nonlinear processes under random quasi phase matching or non-phase matching conditions.

## 4. Experimental section

*BTO thin film growth characterization:* BTO (c-axis) thin film is grown on a dysprosium scandate (DSO, or $DyScO_3$) substrate by pulsed laser deposition with a KrF laser (248 nm wavelength). We first grow a thin layer of 15 nm strontium ruthenate (SRO, or $SrRuO_3$) as bottom electrode at 600 ºC under 100 mTorr oxygen gas pressure with a laser pulse energy density of 2 J·$cm^{-2}$. Then 750 nm BTO thin film was grown epitaxially on top of the SRO layer at 650 ºC under 10 mTorr with a laser pulse energy density of 1.5 J·$cm^{-2}$, followed by cooling at 100 Torr oxygen gas pressure. SRO was exposed at the corners so that electrical contact can later be made.

*Characterization:* The crystallinity of the thin films was characterized by X-ray diffraction (XRD). XRD was performed in the XDD (X-ray Diffractometry and Development) beamline at Singapore Synchrotron Light Source. The X-ray wavelength was 0.15405 nm. The SEM was performed at different accelerating voltages ranging from 5 keV to 30 keV to get the best quality images of the cross-section of the grown BTO/SRO/DSO sample.

For domain analysis, PFM imaging of the sample is conducted prior to poling, examining various regions of the film surface with scanning areas ranging from 5 $\mu$m x 5 $\mu$m to 50 $\mu$m x 50 $\mu$m with drive voltage of 10 V, drive frequency of 15 kHz and lock-in bandwidth of 0.3 kHz using BRUKER Dimension Icon Scanning Probe Microscope (SPM).

*Electrode fabrication:* To fabricate electrodes on the grown BTO film, tungsten was deposited using sputtering system (AJA UHV) with DC power of 200 W for 30 mins which approximately gives the electrode thickness of 100 nm. EBL (Raith) was used to pattern the electrodes with beam current of 5 nA and dose 175 $\mu C/cm^2$. We then used inductively coupled plasma (ICP) etching to etch tungsten and pattern the periodic electrodes.

*Poling experiment:* For optimizing poling voltage and pulse duration, an AWG was used with poling waveform being controlled through our computer using a python code. To amplify the voltage, a HVA (2HVA24-BP1-F-SHV-5KV, UltraVolt, Inc.) was used. The voltage was applied through a poling circuit which uses a voltage divider to measure the poling voltage and current during poling and makes real time adjustments to the HVA to achieve the desired field to realise successful domain inversion.

The applied voltage required for poling was optimized based on the coercive field obtained from the hysteresis analysis of the sample, which came out to be around 50 kV/mm. We started applying voltages slightly higher than this and increased in steps of 5 V to obtain the actual poling voltage. The pulse duration was also increased in the steps of 10 ms, starting from 10 ms.

*PFM imaging:* To visualize the domains, BRUKER Dimension Icon Scanning Probe Microscope (SPM) was used to obtain PFM phase and amplitude images for the poled region. Drive voltage of 10 V, drive frequency of 15 kHz and lock-in bandwidth of 0.3 kHz were used with different scan rates ranging from 0.5 Hz to 1 Hz. The scanned area was varied according to appropriate data sets required for analysis and interpretations.

*Wet etching and SEM imaging:* For selective etching, 0.03 % HF solution was used, and the sample was left in the sample for 150 seconds and 300 seconds. The SEM was then used to obtain the images of the poled region with accelerating voltages of 5 kV. For the tilted images to visualize sidewalls, the tilting angle was kept at 55º.

## Acknowledgement

The authors acknowledge support by the National Research Foundation (NRF), Singapore, under its Competitive Research Program (CRP Award No. NRF-CRP24-2020-0003) and Quantum Engineering Program (QEP-P7).

## Conflict of Interest

The authors declare no conflict of interest.

[1] R. W. Boyd, Nonlinear Optics, 3rd ed., Academic Press, Cambridge, MA, USA **2008**.

[2] A. R. Johnston, J. M. Weingart, *J. Opt. Soc. Am.*, **1965**, *55*, 828.

[3] Robert C Miller, *Phy. Rev.*, **1964,** *134*, , A1313.

[4] G. Arlt; D. Hennings; G. de With, *J. Appl. Phys.,* **1985,** 58, 1619.

[5] R. A. McKee; F. J. Walker; J. R. Conner; E. D. Specht; D. E. Zelmon, *Appl. Phys. Lett.,* **1991,** *59*, 782.

[6] D. H. Kim, H. S. Kwok, *Appl. Phys. Lett.,* **1995,** *67*, 1803–1805

[7] Lucie Mazet, Sang Mo Yang, Sergei,V Kalinin, Sylvie Schamm-Chardon & Catherine Dubourdieu, Science and Tech. of Adv. Mat., **2015**, *16*, 036005.

[8] Keiichi Nashimoto; David K. Fork; Theodore H. Geballe, *Appl. Phys. Lett.,* **1992**, *60*, 1199.

[9] K. J. Kormondy, Y. Popoff, M. Sousa, F. Eltes, D. Caimi, M. D. Rossell, M. Fiebig, P. Hoffmann, C. Marchiori, M. Reinke, M. Trassin, A. A. Demkov, J. Fompeyrine, S. Abel, *Nanotechnology,* **2017,** *28*, 75706.

[10] Stefan Abel, Felix Eltes, J. Elliott Ortmann, Andreas Messner, Pau Castera, Tino Wagner, Darius Urbonas, Alvaro Rosa, Ana M. Gutierrez, Domenico Tulli, Ping Ma, Benedikt Baeuerle, Arne Josten , Wolfgang Heni 3, Daniele Caimi1, Lukas Czornomaz, Alexander A. Demkov, Juerg Leuthold, Pablo Sanchis and Jean Fompeyrine, *Nat. Mat.*, **2019**, *18*, 42.

[11] Artemios Karvounis, Flavia Timpu, Viola V. Vogler-Neuling, Romolo Savo, and Rachel Grange, Adv. Optical Mater., **2020,** *8*, 2001249.

[12] F. Leroy, A. Rousseau, S. Payan, E. Dogheche, D. Jenkins, D. Decoster, M. Maglione, Opt. Lett., **2013**, *38*, 1037.

[13] Yu Cao, Siew Li Tan, Eric Jun Hao Cheung, Shawn Yohanes Siew, Changjian Li, Yan Liu, Chi Sin Tang, Manohar Lal, Guanyu Chen, Karim Dogheche, Ping Yang, Steven Pennycook, Andrew Thye Shen Wee, Soojin Chua, Elhadj Dogheche, Thirumalai Venkatesan, and Aaron Danner, Adv. Mater. , **2021** *33*, 2101128.

[14] A. Petraru, J. Schubert, M. Schmid, Ch. Buchal, *Appl. Phys. Lett.*, **2002**, *81,* 1375.

[15] Pingsheng Tang, D. J. Towner, A. L. Meier and B. W. Wessels, *IEEE Phot. Tech. Lett.* , **2004,** *16*, 1837.

[16] P Tang, D. J. Towner, T. Hamano, A. L. Meier and B. W. Wessels, *Opt. Exp.* , **2004,** *12*, 5962.

[17] A. Petraru, J. Schubert, M. Schmid, O. Trithaveesak, and Ch. Buchal, *Opt. Lett*., **2003,** *28*, 2527

[18] Na Sun, DeGui Sun, Di Wu, Yinghui Guo, Yulong Fan, Fang Zou, Mingbo Pu, and Xiangang Luo, *Laser Photonics Rev*., **2023,** 2300937 (1-9).

[19] I.-D. Kim, Y. Avrahami, L. Socci, F. Lopez-Royo, H. L. Tuller, *J. Asian Ceram. Soc*, **2014**.*, 2*, 231.

[20] D. M. Gill, C. W. Conrad, G. Ford, B. W. Wessels, and S. T. Ho, Appl. Phys. Lett., **1997**, *71*, 1783.

[21] R. W. Boyd, Nonlinear Optics, 3rd ed., Academic Press, Cambridge, MA, USA **2008.**

[22] Valdas Pasiskevicius, Gustav Strömqvist, Fredrik Laurell, Carlota Canalias, Opt. Mat. , **2012,** *34*, 513.

[23] M Houe and P D Townsend, *J. Phys. D: Appl. Phys.*, **1995***, 28*, 1747.

[24] Juanjuan Lu, Ayed Al Sayem, Zheng Gong, Joshua B. Surya, Chang-Ling Zou, and Hong X. Tang, Optica, **2021,** *8*, 539.

[25] Milad Gholipour Vazimali, Sasan Fathpour, Adv. Phot., **2022,** *4*, 034001.

[26] Ayed Al Sayem, Yubo Wang, Juanjuan Lu, Xianwen Liu, Alexander W. Bruch, and Hong X. Tang, *Appl. Phys. Lett.,* **2021**, *119*, 231104.

[27] Juanjuan Lu, Ming Li, Chang-Ling Zou, Ayed Al Sayem, and Hong X. Tang, Optica, **2020**, *7*, 1654.

[28] S. D. Setzler, P. G. Schunemann, T. M. Pollak, L. A. Pomeranz, M. J. Missey, and David E. Zelmon, *Advanced Solid State Lasers*, **1999,** paper MD.

[29] J. Zhou, M. Liu, C. Zhou, B. H. Kuo, F. -S. Shieu and P. T. Lin, *IEEE Journal of Selected Topics in Quantum Electronics*, **2023,** *29*, 5100606(1-6).

[30] Pao Tai Lin; B. W. Wessels; Joon I. Jang; J. B. Ketterson, *Appl. Phys. Lett.*, **2008**, *92*, 221103.

[31] Junchao Zhou, Wenrui Zhang, Mingzhao Liu, and Pao Tai Lin, Photon. Res., **2019,** *7*, 1193.

[32] Eugene Kim, Andrea Steinbruck Maria Teresa Buscaglia, Vincenzo Buscaglia, Thomas Pertsch, and Rachel Grange, *ACS Nano*, **2013**, *7*, 5343.

[33] Bipin Bihari; Jayant Kumar; Gregory T. Stauf; Peter C. Van Buskirk; Cheol Seong Hwang, *J. Appl. Phys.,* **1994,** *76*, 1169.

[34] Ülle-Linda Talts, Helena C. Weigand, Grégoire Saerens, Peter Benedek, Joel Winiger, Vanessa Wood, Jürg Leuthold, Viola Vogler-Neuling, and Rachel Grange, *Small,* **2023,** *19*, 2304355 (1-7).

[35] Yu Cao, Nour Al Meselmene, Elhadj Dogheche, Ping Yang, Parikshit Moitra, Shi Qiang Li, Thirumalai Venkatesan, and Aaron Danner, Opt. Mater. Exp., **2023,** *13*, 152.

[36] Yu Cao, Jun Da Ng, Hong-Lin Lin, Siew Li Tan, Aaron Danner, *Appl. Phy. Lett.,* **2023,** *122*, 031106.

[37] Yu Cao, Haidong Liang, Hong-Lin Lin, Luo Qi, Ping Yang, Xuanyao Fong, Elhadj Dogheche, Andrew Bettiol, and Aaron Danner, Nano Letters, **2023,** *23* (16), 7267.

[38] Licheng Ge, Yuping Chen, Haowei Jiang, Guangzhen Li, Bing Zhu, Yi'an Liu, and Xianfeng Chen, , **2018,** *6*, 954.